\documentclass{tlp}
\usepackage{code}
\usepackage{url}

\title{Logic Programming with Satisfiability}

\author[M. Codish, V. Lagoon, and P.J. Stuckey]{
MICHAEL CODISH \\
Department of Computer Science, Ben-Gurion University, Israel 
\email{mcodish@cs.bgu.ac.il}
\and
VITALY LAGOON \\
 Department of Computer Science \& Software Engineering,
  University of Melbourne, Australia \\
\email{vitaly.lagoon@gmail.com}
\and
PETER J. STUCKEY \\
NICTA Victoria Laboratory and  Department of Computer Science \& Software Engineering, \\
  University of Melbourne, Australia
\email{pjs@csse.unimelb.edu.au}
}

\submitted{10 May 2006}
\accepted{12 February 2007}
\revised{30 January 2007}

\def\etal{{\it et al{.}}}

\begin{document}
\maketitle

\begin{abstract}
\begin{sloppypar}
  This paper presents a Prolog interface to the MiniSat satisfiability
  solver. Logic programming with satisfiability combines the strengths
  of the two paradigms: logic programming for encoding search problems
  into satisfiability on the one hand and efficient SAT solving on the
  other. This synergy between these two exposes a programming paradigm
  which we propose here as a logic programming pearl.  To illustrate
  logic programming with SAT solving we give an example Prolog program
  which solves instances of Partial MAXSAT.

To appear in Theory and 
Practice of Logic Programming (TPLP)
\end{sloppypar}
\end{abstract}

\section{Introduction}

The use of SAT solvers in a wide range of applications is a great
success of recent years and quickly growing more popular.
Contributing to this success are two main factors: (a) SAT solvers are
continuously becoming more powerful, and (b) encoding techniques are
emerging to represent a wide range of search problems as propositional
formulae such that each satisfying assignment of the encoding
represents a solution of the problem.

This paper presents a Prolog interface to the MiniSat SAT solver
\cite{EenS03,minisat-page}. MiniSat is a small ($\approx 1200$ lines
of C-code) and efficient open-source SAT solver which is designed to
enable easy integration with other tools and languages.  Application
of this interface facilitates the best of both worlds: Prolog
programming for encoding a search problem into a propositional formula
on the one hand, and application of powerful SAT solving tools on the
other.

In recent work \cite{RTA2006} we apply SAT encodings to decide LPO
termination \cite{Dershowitz82} of a given term rewrite system $\tau$.
Here LPO termination of $\tau$ is encoded as a propositional formula
$\varphi_\tau$ and any assignment satisfying $\varphi_\tau$ indicates
a precedence on the symbols in $\tau$ such that its corresponding
lexicographic path order orients the rules of $\tau$. Namely for each
rule $\ell\rightarrow r$ in $\tau$, $\ell\succ_{lpo} r$ holds.  Prolog
is well suited for the tasks of parsing the given system $\tau$ and
constructing the encoding $\varphi_\tau$. Searching for an assignment
which satisfies $\varphi_\tau$ is conveniently delegated to the SAT
solver. 

In other recent works Codish \etal\ apply SAT encodings to determine
termination of rewrite systems using dependency pairs \cite{ArtsG00}
and using recursive path orders (RPO) \cite{Dershowitz82}.  A
contribution of those works is the propositional encodings for so
called \emph{argument filterings} and \emph{usable rules}, in
\cite{LPAR2006}, and the encodings for multi-set orders and
lexicographic path orders modulo permutation of arguments, in
\cite{LPAR2006:short}.

All of the above mentioned results have been developed on top of the
described Prolog MiniSat interface. We have also had positive
experience in using the interface as an educational tool for teaching
logic programming concepts as well as the basics of SAT encodings and
SAT solving.

\section{Preliminaries}

Most SAT solvers assume as input a propositional formula in
conjunctive normal form (CNF). That is a conjunction of disjunctions
of literals, or equivalently a conjunction of clauses. Each literal is
a propositional variable $p$ or its negation $\neg p$. A truth
assignment is a mapping from propositional variables into $\{0,1\}$.

\paragraph{\bf Syntax:~}
In the logic programming setting we represent: literals as terms of
the form \texttt{X} or\texttt{ -X} where \texttt{X} is a logic
variable; clauses as lists of literals; and conjunctions of clauses as
lists of clauses.  For propositional formulae we use terms involving
the symbols \verb!0/0!, \verb!1/0!, \verb!-/1!, \verb!*/2!,
\verb!+/2!, \verb!==/2!, \verb!xor/2! for: false, true, negation,
conjunction, disjunction, bi-implication and xor.  The syntax supports
also: \texttt{(X->Y;Z)} which is equivalent to \texttt{X*Y+(-X)*Z}.
In the logic programming setting a truth assignment is a substitution
of the variables to the constants $\{0,1\}$.
We also use a list of propositions to represent non-negative integers,
in two different ways.  The unary representation is
given by the sum of the bits. For example [0,1,0,1] represents 2.
The binary representation is as usual, 
but giving the least significant bit first. For example [0,1,0,1] 
represents $0 \times 2^0 + 1 \times 2^1 + 0 \times 2^2 + 1 \times 2^3 = 10$.

\medskip We now recall the satisfiability and maximum satisfiability
problems.  Partial MAXSAT is a generalization of SAT and MAXSAT
introduced by Miyazaki and Iwama \cite{MiyazakiIK96}. See also
\cite{ChaIKM97}.

\paragraph{\bf SAT:~}Given a propositional logic formula $\varphi$ in
conjunctive normal form, is there an assignment of truth values to the
propositional variables that makes $\varphi$ true.
\paragraph{\bf MAXSAT:~}Given a propositional logic formula $\varphi$
in conjunctive normal form, find an assignment of the logical
variables that maximizes the number of clauses in $\varphi$ that are
true.

\paragraph{\bf Partial MAXSAT:~}Given propositional logic formulae
$\varphi$ and $\psi$ in conjunctive normal form, the problem is to
find an assignment that satisfies all clauses in $\varphi$ and
maximizes the number of clauses in $\psi$ that are true.

\section{Interfacing Prolog with MiniSat}

The library implementation is written primarily in SWI-Prolog
\cite{swi-prolog,swi-page} and interfaces the MiniSat solver
\cite{EenS03,minisat-page} for solving SAT instances.  We have
integrated MiniSat and SWI-Prolog through $\approx$190 lines of C-code
and $\approx$140 lines of Prolog code.  The C-code handles the
low-level interface and conversion between the internal data
representations of SWI-Prolog and of MiniSat. The Prolog code provides
a high-level interface for using the SAT solver in Prolog applications.
The SAT solver is deterministic. It does not, and is not intended to,
provide alternative satisfying assignments over backtracking.
The MiniSat library for Prolog can be downloaded from
\cite{minisat-prolog}. It includes three modules for the user:
\texttt{cnf.pl}, \texttt{adder.pl} and \texttt{minisat.pl}.

The module \texttt{cnf.pl} exports the predicate \texttt{cnf/2}.  A
call to \texttt{cnf(F, Cnf)} transforms a propositional formula
\texttt{F} to a conjunctive normal form \texttt{Cnf}.
The implementation of \texttt{cnf(F,Cnf)} applies a Tseitin
transformation~\cite{tseitin} to ensure that the size of the
conjunctive normal form \texttt{Cnf} is linear in that of the input
formula \texttt{F}.
The basic idea is to associate fresh variables with the sub-formulae
of \texttt{F}. So, for example, if \texttt{F} is of the form
\texttt{G+H} and variables \texttt{A, B, C} are associated
respectively with sub-formulae \texttt{F, G, H}, then the clauses for
the bi-implication \texttt{A $\leftrightarrow$ (B $\lor$ C)} are
introduced to the \texttt{Cnf} together with the clauses for the
transformations of \texttt{B $\leftrightarrow$ G} and \texttt{C
  $\leftrightarrow$ H}. This process establishes the conjunctive
normal form of \texttt{A $\leftrightarrow$ F}. Adding the singleton
clause \texttt{[A]} then provides a CNF equisatisfiable to $F$.  The
variables \texttt{A, B, C} are sometimes referred to as ``Tseitin
variables''.

A technique proposed by Plaisted and Greenbaum \cite{PG86} is applied
to reduce the size of the transformation. If it can be determined that
we are only interested in the truth of a Tseitin variable, we can
simplify the transformation only including the implication.  In the
above example, since we are only interested in the truth of Tseitin
variable \texttt{A} (for \texttt{F}), the CNF created for \texttt{A
  $\leftrightarrow$ (B $\lor$ C)}, will be \texttt{[[-A,B,C]]} rather
than \texttt{[[-A,B,C],[A,-B],[A,-C]]}. Similarly when we are only
interested in falsity of a Tseitin variable.

To demonstrate the use of the module consider the following queries
where \texttt{T}, \texttt{T1} and \texttt{T2} are the Tseitin
variables.

\begin{code}
   ?- cnf(X==Y,Cnf).
   Cnf = [[T], [-X, Y, -T], [X, -Y, -T]] 
   ?- cnf((X*Y)+(-X*Z),Cnf).
   Cnf = [[T], [-T, T1, T2], [-T2, -X], 
          [-T2, Z], [-T1, X], [-T1, Y]] 
\end{code} 

Figure~\ref{fig:adder} illustrates the module \texttt{adder.pl} which
includes a textbook \cite{CLR:1990} ripple-carry
circuit for binary addition.
The module exports the predicate \texttt{sum/3}. A call to
\texttt{sum(Unary, Binary, Psi)} constructs a Boolean circuit
\texttt{Psi}. The arguments \texttt{Unary} and \texttt{Binary} are
lists of Boolean formulae, the truth values of which encode unary and
binary numbers respectively. 
These are the inputs and outputs of the
circuit which can be seen as computing the binary sum of its unary
inputs.
The formula, \texttt{Psi} is a propositional statement capturing the
correspondence between unary and binary representations of
non-negative integers.  The formula \texttt{Psi} is satisfiable
exactly when \texttt{Unary} and \texttt{Binary} encode the same
number.
In the recursive call of the code, the formula \texttt{Psi=F1*F2*F3}
is constructed by splitting the inputs into two equal halves (padding
with a zero if necessary). From the recursive calls, the subformulae
\texttt{F1} and \texttt{F2} relate the two halves of the inputs
(unary) to the (binary) numbers \texttt{S1} and \texttt{S2}.  The call
to \texttt{add(S1,S2,Sum,F3)} constructs the formula \texttt{F3} for
the binary \texttt{Sum} of \texttt{S1} and \texttt{S2}.
Observe that the length of \texttt{Binary} is $\lceil \log_2
|\mathtt{Unary}| \rceil$.
% and that \texttt{Psi} is LINEAR in the size
% of its inputs.   mc: i think this is NOT TRUE. each adder is linear
% in size of its inputs but there are log n levels of adders at each
% level they have total of n elements

\begin{figure}[hbtp]
  \centering
  \begin{code}
  :- module(adder,[sum/3]).

  % sum(+,-,-).
  sum([B],[S],(S==B)).
  sum([B1,B2|Bs],Sum,F1*F2*F3) :-
          split([B1,B2|Bs],Xs,Ys),
          sum(Xs,S1,F1), sum(Ys,S2,F2), add(S1,S2,Sum,F3).

  % split(+,-,-).
  split([],[],[]).
  split([X],[X],[0]).
  split([X,Y|XYs],[X|Xs],[Y|Ys]) :- split(XYs,Xs,Ys).

  % add(+,+,-,-).
  add([X|Xs],[Y|Ys],[Z|Zs],(CXY==CarryXY)*(Z==SumXY)*Sum) :-
          halfadder(X,Y,SumXY,CarryXY),
          adder(Xs,Ys,CXY,Zs,Sum).

  % adder(+,+,-,-).
  adder([],[],Carry,[Z],Z==Carry).
  adder([X|Xs],[Y|Ys],Carry,[Z|Zs],(CXY==CarryXY)*(Z==SumXY)*Rest) :-
          fulladder(X,Y,Carry,SumXY,CarryXY),
          adder(Xs,Ys,CXY,Zs,Rest).

  % fulladder(+,+,+,-,-).
  fulladder(X, Y, C, (X xor Y xor C), (C -> X+Y ; X*Y) ).
  % halfadder(+,+,-,-).
  halfadder(X, Y,    (X xor Y),       X*Y              ).
  \end{code}
  \caption{A ripple-carry adder circuit}
  \label{fig:adder}
\end{figure}

To demonstrate the use of the module consider the following query
constructing the circuit \texttt{Psi} for the inputs \texttt{Unary =
  [X+Y,X*Y,X==Y,X xor Y]} and outputs \texttt{Binary}.
  \begin{code}
   ?- sum([X+Y,X*Y,X==Y,X xor Y],Binary,Psi).
   Binary = [S1, S2, S3]
   Psi = (T1==X+Y)*(T2==(X==Y))*(T3==T1 xor T2)*(T4==T1*T2)*(T5==X*Y)* 
      (T6==X xor Y)*(T7==T5 xor T6)* (T8==T5*T6)*(S1==T3 xor T7)*
      (S2==T4 xor T8 xor (T3*T7))* (S3==(T3*T7->T4+T8;T4*T8)) 
  \end{code}
  Table~\ref{table:sum} illustrates the declarative meaning of the
  predicate depicting the truth values for \texttt{Unary} and
  \texttt{Binary} determined by \texttt{Psi} for the 4 truth
  assignments of \texttt{X} and \texttt{Y}.

  \begin{table}[hbtp]
    \centering
    \begin{tabular}{|l|l|l|l|}
\cline{1-4}
X  & Y  & Unary    &          Binary \\
\cline{1-4}
0   & 0   & [0, 0, 1, 0]   & [1, 0, 0] \\
0   & 1   & [1, 0, 0, 1]   & [0, 1, 0] \\
1   & 0   & [1, 0, 0, 1]   & [0, 1, 0] \\
1   & 1   & [1, 1, 1, 0]   & [1, 1, 0] \\
\cline{1-4}
    \end{tabular}
    \caption{Truth values for the circuit
      summing \texttt{Unary=[X+Y,X*Y,X==Y,X xor Y]}}
    \label{table:sum}
  \end{table}

The module \texttt{minisat.pl} exports four predicates:
\begin{itemize}
\item \texttt{solve(Cnf)} succeeds if and only if the formula
  \texttt{Cnf} in conjunctive normal form is satisfiable, binding its
  variables to truth values 0 (false) and 1 (true) that satisfy 
\texttt{Cnf}.
\item \texttt{sat(Cnf)} succeeds if and only if the formula
  \texttt{Cnf} in conjunctive normal form is satisfiable. It is
  similar to \texttt{solve(Cnf)} but does
  not bind any variables.

\item %% \pjs{modified description}
\texttt{minimize(Vec,Cnf)} is similar to \texttt{sat(Cnf)}. The
  additional argument \texttt{Vec} is a list of variables (e.g.,
  occurring in \texttt{Cnf}).  
  The variables of \texttt{Vec} are assigned the truth value
  that minimizes the binary number represented by \texttt{Vec}
  for all solutions of \texttt{Cnf}. If \texttt{Cnf}
  has no solutions (i.e. is unsatisfiable) it fails.

%% \item \texttt{minimize(Vec,Cnf)} is similar to \texttt{sat(Cnf)}. The
%%   additional argument \texttt{Vec} is a list of variables (e.g.,
%%   occurring in \texttt{Cnf}).  Here, if \texttt{Cnf} is satisfiable,
%%   these variables are assigned truth values minimizing the binary
%%   number represented by \texttt{Vec}.

\item \texttt{maximize(Vec,Cnf)} is similar to
  \texttt{minimize(Vec,Cnf)} but the assignment returned maximizes the
  value of the non-negative integer represented by \texttt{Vec}.
\end{itemize}

The predicates \texttt{minimize/2} and \texttt{maximize/2} are
illustrated in Figure~\ref{fig:minmax}. Consider the query \texttt{?-
  maximize(Xs, Cnf)} where \texttt{Xs} is a list of $k$ variables and
\texttt{Cnf} a formula in conjunctive normal form.
To solve the query, the basic idea is to pose $k$ questions, one for
each variable of \texttt{Xs}, to the SAT solver to determine the
maximum value of \texttt{Xs}. Each question determines the
satisfiability of \texttt{Cnf} when setting the next bit in \texttt{Xs}
to its maximal value.  Observe, in the code, that these questions are
posed using \texttt{sat} so as not to bind variables in the formula.
Note also that each of these questions is of size $O(|\mathtt{Cnf}|)$.

%% \pjs{Added base case call to sat, for the case that CNF is
%% unsatisfiable the whole thing fails}
\begin{figure}[hbtp]
  \centering
  \begin{code}
% minimize(+,+).
minimize([],CNF) :- sat(CNF).
minimize([B|Bs],CNF) :- minimize(Bs,CNF), ( (B=0, sat(CNF)) -> true ; B=1 ).

% maximize(+,+).
maximize([],CNF) :- sat(CNF).
maximize([B|Bs],CNF) :- maximize(Bs,CNF), ( (B=1, sat(CNF)) -> true ; B=0 ).

  \end{code}
  \caption{Minimization and maximization in \texttt{minisat.pl}}
  \label{fig:minmax}
\end{figure}

To demonstrate the use of the module consider the following queries:
\begin{itemize}
\item \texttt{solve/1}. The call succeeds binding \texttt{X} and
  \texttt{Y} to truth values.
  \begin{code}
 ?- cnf(X==Y,Cnf), solve(Cnf).
 X=0, Y=0
  \end{code}\vspace{-5mm}
\item \texttt{sat/1}. The call succeeds without binding \texttt{X} and
  \texttt{Y}.
  \begin{code}
 ?- cnf(X==Y,Cnf), sat(Cnf).
 Yes
  \end{code}\vspace{-5mm} 
\item \texttt{sum/3} with \texttt{solve/1}. The first call succeeds binding
  \texttt{X=0} and \texttt{Y=0}. In this case the
  circuit output is 1 (only one of the inputs is true under this
  assignment). The second call indicates that it is possible to
  satisfy 2 of the inputs under the assignment  \texttt{X=0} and
  \texttt{Y=1}.
  \begin{code}
 ?- sum([X+Y,X*Y,X==Y,X xor Y],Sum,F), cnf(F,Cnf), solve(Cnf).
 X = 0, Y = 0
 Sum = [1, 0, 0]
 ?- sum([X+Y,X*Y,X==Y,X xor Y],[0,1,0],F), cnf(F,Cnf), solve(Cnf).
 X = 0, Y = 1
  \end{code}\vspace{-5mm} 
\item \texttt{sum/3} with \texttt{maximize/2}. The answer
  \texttt{Sum=[1,1,0]} indicates that it is possible to satisfy at
  most three of the four formulae.  The call
  \texttt{maximize(Sum,Cnf)} binds only the elements of \texttt{Sum}.
  To obtain the maximizing assignment, \texttt{X=1, Y=1}, the call to
  \texttt{maximize} must be followed by a call to \texttt{solve(Cnf)}.
  \begin{code}
 ?- sum([X+Y,X*Y,X==Y,X xor Y],Sum,F), cnf(F,Cnf),
    maximize(Sum,Cnf), solve(Cnf).
  Sum=[1,1,0]
  X=1, Y=1
  \end{code}
\end{itemize}

% ?- sum([X+Y,X*Y,X==Y,X xor Y],Sum,F),cnf(F,Cnf),maximize(Sum,Cnf).

% 

\section{Solving Partial MAX-SAT}

To solve a Partial MAXSAT problem for CNF formula $\varphi$ and $\psi$
we seek an assignment that satisfies $\varphi$ and maximizes the
number of clauses of $\psi$ which are satisfied. We solve the more
general problem with $\varphi$ an arbitrary propositional formula and
$\psi$ a list of $n$ propositional formulae. 

The solution is illustrated in Figure~\ref{fig:pmaxsat}. The arguments
\texttt{Phi} and \texttt{Psi} correspond to $\varphi$ and $\psi$
respectively.  The key idea is to construct the formula
\texttt{SumPsi} representing the sum of the $n$ formulae in
\texttt{Psi}. This results in a vector \texttt{Max} with $\lceil \log_2 n\rceil$
bits.  We then aim to satisfy the conjunction \texttt{Phi*SumPsi}
while maximizing the  number represented by \texttt{Max}.
% The size of \texttt{SumPsi} is linear in the sum of the sizes of the
% $n$ formulae of \texttt{Psi}.  
Solving the query \texttt{maximize(Max,Cnf)} involves $O(\log n)$ calls
to the SAT solver.
 
\begin{figure}[hbtp]
  \centering
\begin{code}
   % partialMaxSat(+,+).
   partialMaxSat(Phi,Psi) :-
         sum(Psi,Max,SumPsi), cnf(Phi*SumPsi,Cnf), 
         maximize(Max,Cnf),   solve(Cnf).
\end{code}
  \caption{Solving Partial MAXSAT }
  \label{fig:pmaxsat}
\end{figure}

For example the following query provides an assignment which satisfies 
$\varphi=X+Y$ and 4 of the 7 formula in the second argument.
\begin{code}
   ?- partialMaxSat(X+Y,[X*Y,X==Y,X xor Y,-X+Y, -X, -Y, X]).
   X = 1, Y = 1
\end{code}

\section{Conclusion}

We define a Prolog library for solving SAT instances through an
interface to the MiniSat solver. 
The combination of Prolog for manipulating Boolean formulae, 
and powerful SAT tools for solving them is compelling.
We can straightforwardly build solutions to difficult problems, 
with small and clean, pearlish code.
We illustrate this in the paper by encoding a
reduction from Partial MAXSAT to SAT. 
We have successfully used the library to solve many such 
problems~\cite{RTA2006,LPAR2006,LPAR2006:short}.

%\bibliography{paper}

\begin{thebibliography}{}

\bibitem[\protect\citeauthoryear{Annov, Codish, Giesl, Schneider-Kamp, and
  Thiemann}{Annov et~al\mbox{.}}{2006}]{LPAR2006:short}
{\sc Annov, E.}, {\sc Codish, M.}, {\sc Giesl, J.}, {\sc Schneider-Kamp, P.},
  {\sc and} {\sc Thiemann, R.} 2006.
\newblock A sat-based implementation for {RPO} termination.
\newblock \url{http://www.lix.polytechnique.fr/~hermann/LPAR2006/short.html}.
\newblock Short Paper at LPAR.

\bibitem[\protect\citeauthoryear{Arts and Giesl}{Arts and
  Giesl}{2000}]{ArtsG00}
{\sc Arts, T.} {\sc and} {\sc Giesl, J.} 2000.
\newblock Termination of term rewriting using dependency pairs.
\newblock {\em Theoretical Computer Science\/}~{\em 236,\/}~1-2, 133--178.

\bibitem[\protect\citeauthoryear{Cha, Iwama, Kambayashi, and Miyazaki}{Cha
  et~al\mbox{.}}{1997}]{ChaIKM97}
{\sc Cha, B.}, {\sc Iwama, K.}, {\sc Kambayashi, Y.}, {\sc and} {\sc Miyazaki,
  S.} 1997.
\newblock Local search algorithms for partial {MAXSAT}.
\newblock In {\em AAAI/IAAI}. 263--268.

\bibitem[\protect\citeauthoryear{Codish, Lagoon, and Stuckey}{Codish
  et~al\mbox{.}}{2006a}]{minisat-prolog}
{\sc Codish, M.}, {\sc Lagoon, V.}, {\sc and} {\sc Stuckey, P.~J.} 2006a.
\newblock Minisat library for prolog.
\newblock \url{http://www.cs.bgu.ac.il/~mcodish/Software/pl-minisat.tgz}.

\bibitem[\protect\citeauthoryear{Codish, Lagoon, and Stuckey}{Codish
  et~al\mbox{.}}{2006b}]{RTA2006}
{\sc Codish, M.}, {\sc Lagoon, V.}, {\sc and} {\sc Stuckey, P.~J.} 2006b.
\newblock Solving partial order constraints for {LPO} termination.
\newblock In {\em Term Rewriting and Applications (RTA)}. Vol. 4098. Springer,
  Seattle, USA, 4--18.

\bibitem[\protect\citeauthoryear{Codish, Schneider-Kamp, Lagoon, Thiemann, and
  Giesl}{Codish et~al\mbox{.}}{2006}]{LPAR2006}
{\sc Codish, M.}, {\sc Schneider-Kamp, P.}, {\sc Lagoon, V.}, {\sc Thiemann,
  R.}, {\sc and} {\sc Giesl, J.} 2006.
\newblock Sat solving for argument filterings.
\newblock In {\em Logic for Programming, Artificial Intelligence and Reasoning
  (LPAR)}. Vol. 4246. Springer, 30--44.

\bibitem[\protect\citeauthoryear{Cormen, Leiserson, and Rivest}{Cormen
  et~al\mbox{.}}{1990}]{CLR:1990}
{\sc Cormen, T.~H.}, {\sc Leiserson, C.~E.}, {\sc and} {\sc Rivest, R.~L.}
  1990.
\newblock {\em Introduction to Algorithms}.
\newblock MIT Press, Chapter~29.

\bibitem[\protect\citeauthoryear{Dershowitz}{Dershowitz}{1982}]{Dershowitz82}
{\sc Dershowitz, N.} 1982.
\newblock Orderings for term-rewriting systems.
\newblock {\em Theoretical Computer Science\/}~{\em 17}, 279--301.

\bibitem[\protect\citeauthoryear{E{\'e}n and S{\"o}rensson}{E{\'e}n and
  S{\"o}rensson}{2004}]{EenS03}
{\sc E{\'e}n, N.} {\sc and} {\sc S{\"o}rensson, N.} 2004.
\newblock An extensible {SAT}-solver.
\newblock In {\em Theory and Applications of Satisfiability Testing, 6th
  International Conference, SAT 2003 (Selected Revised Papers)},
  {E.~Giunchiglia} {and} {A.~Tacchella}, Eds. Lecture Notes in Computer
  Science, vol. 2919. Springer, 502--518.

\bibitem[\protect\citeauthoryear{??}{MiniSAT}{2006}]{minisat-page}
MiniSAT 2006.
\newblock Mini{SAT} solver.
\newblock \url{http://www.cs.chalmers.se/Cs/Research/FormalMethods/MiniSat}.
\newblock Viewed December 2005.

\bibitem[\protect\citeauthoryear{Miyazaki, Iwama, and Kambayashi}{Miyazaki
  et~al\mbox{.}}{1996}]{MiyazakiIK96}
{\sc Miyazaki, S.}, {\sc Iwama, K.}, {\sc and} {\sc Kambayashi, Y.} 1996.
\newblock Database queries as combinatorial optimization problems.
\newblock In {\em CODAS}. 477--483.

\bibitem[\protect\citeauthoryear{Plaisted and Greenbaum}{Plaisted and
  Greenbaum}{1986}]{PG86}
{\sc Plaisted, D.} {\sc and} {\sc Greenbaum, S.} 1986.
\newblock A structure preserving clause form translation.
\newblock {\em Journal of Symbolic Computation\/}~{\em 2}, 293--304.

\bibitem[\protect\citeauthoryear{??}{SWI-Prolog}{}]{swi-page}
SWI-Prolog.
\newblock {SWI}-prolog.
\newblock \url{http://http://www.swi-prolog.org/}.
\newblock Viewed December 2005.

\bibitem[\protect\citeauthoryear{Tseitin}{Tseitin}{1968}]{tseitin}
{\sc Tseitin, G.} 1968.
\newblock On the complexity of derivation in propositional calculus.
\newblock In {\em Studies in Constructive Mathematics and Mathematical Logic}.
  115--125.
\newblock Reprinted in J. Siekmann and G. Wrightson (editors), Automation of
  Reasoning, vol. 2, pp. 466-483, Springer-Verlag Berlin, 1983.

\bibitem[\protect\citeauthoryear{Wielemaker}{Wielemaker}{2003}]{swi-prolog}
{\sc Wielemaker, J.} 2003.
\newblock An overview of the {SWI-Prolog} programming environment.
\newblock In {\em Proceedings of the 13th International Workshop on Logic
  Programming Environments}, {F.~Mesnard} {and} {A.~Serebenik}, Eds. Katholieke
  Universiteit Leuven, Heverlee, Belgium, 1--16.
\newblock CW 371.

\end{thebibliography}
%\bibliographystyle{acmtrans}

\end{document}